\documentclass[12pt,preprint]{aastex}

\shorttitle{Recnnection Inflow and Outflow}
\shortauthors{Takasao et al.}

\begin{document}

\title{Simultaneous Observation of Reconnection Inflow and Outflow Associated with the 2010 August 18 Solar Flare}

\author{Shinsuke Takasao\altaffilmark{1,2}, 
Ayumi Asai\altaffilmark{3}, 
Hiroaki Isobe\altaffilmark{3}, and 
Kazunari Shibata\altaffilmark{1}}

\email{takasao@kwasan.kyoto-u.ac.jp}
\altaffiltext{1}{Kwasan and Hida Observatories, Kyoto University, 
Yamashina, Kyoto 607-8471, Japan}
\altaffiltext{2}{Department of Astronomy, Kyoto University, 
Sakyo, Kyoto 606-8502, Japan}
\altaffiltext{3}{Unit of Synergetic Studies for Space, Kyoto University,
Yamashina, Kyoto 607-8471, Japan}

\begin{abstract}
We report the simultaneous extreme ultraviolet observation of magnetic
reconnection inflow and outflow in a flare on 2010 August 18 observed by the
Atmospheric Imaging Assembly on board the {\it Solar Dynamics
Observatory}.
We found that during the rise phase of the flare, some plasma blobs appeared in
the sheet structure above the hot loops.  The plasma blobs
were ejected bidirectionally along the sheet structure (outflow), at the same time as
the threads visible in extreme ultraviolet images moved toward the sheet structure (inflow).  The
upward and downward ejection velocities are 220-460~km~s$^{-1}$ and
250-280~km~s$^{-1}$, respectively.  The inflow speed changed from
90~km~s$^{-1}$ to 12~km~s$^{-1}$ in 5 minutes.
By using these velocities, we estimated the nondimensional reconnection rate,
which we found to vary during this period from 0.20 to 0.055.
We also found that  the plasma blobs in the sheet structure 
collided or merged with each other before they were ejected
from the sheet structure.
We hypothesize that the sheet structure is the current sheet and that these plasma blobs are plasmoids or magnetic islands, which could be important for understanding the dynamics of the reconnection region.
\end{abstract}

\keywords{magnetic reconnection --- Sun: corona --- Sun: flares --- 
Sun: magnetic field --- Sun: UV radiation}

\section{Introduction}

Solar flares are the most energetic phenomenon in the solar atmosphere.
Huge amounts of magnetic energy ($10^{29}$\,-\,$10^{32}$\,erg), that has been stored in the corona, is released over a short period of time ($10^2$\,-\,$10^4$~sec).
It is widely believed that magnetic reconnection, a physical process in which a magnetic field in a highly conducting plasma changes its connectivity due to finite resistivity, plays a central role in the rapid dissipation of magnetic energy in solar flares, as well as other explosive phenomena in astrophysical, space, and laboratory plasmas \citep{yam10}.

Many studies have reported the observational features of magnetic reconnection in solar flares.
These studies include observations of cusp-shaped loops \citep{tsu92}, loop-top hard X-ray sources \citep{mas94}, plasmoid ejection \citep{shi95,tsu97,ohy98},
reconnection inflows \citep{yok01,nar06,li09}, reconnection outflows \citep{wan07}, supra-arcade downflows \citep{mc99,inn03,asa04,sav11}
and current-sheet-like structures \citep{sui03,lin05,liu10}. 
Yokoyama et al.(2001) found a pattern of bright threads that merged toward an X-point (where the magnetic field was hypothesized to be reconnecting) in extreme ultraviolet (EUV) images, and identified the motion as reconnection inflow. 
Narukage \& Shibata (2006) further worked on the statistical study of inflow by EUV observations, and found that the observed inflows are closely associated with plasmoid ejections or coronal mass ejections.

The classic magnetic reconnection model for solar flares, the CSHKP model \citep{car64,stu66,hir74,kp76}, successfully explains some features of the long-lived (more than 1 hour) flare. 
This model has been improved to include a generalized configuration, leading to a unified model proposed by Shibata et al. (1995). 
Many of the observations presented above relating to the global structure of flares have been successfully explained by this magnetic reconnection model.
For reviews of our current knowledge of magnetic reconnection in solar flares see, for example, Shibata (1999).

There are still fundamental difficulties in the theory of magnetic reconnection in flares, especially on 
the physics that determine the reconnection rate. The reconnection rate is defined as the reconnected magnetic flux per unit time, where 
the dimensionless form is the Alfv\'en Mach number $M_{A}=V_{\rm inflow}/V_{A}$, where $V_{\rm inflow}$ is the inflow speed into the reconnection point and $V_{A}$ is the Alfv\'en speed, respectively.
Observationally the value of the reconnection rate is estimated to be 0.001 - 0.1 \citep{iso02,iso05,nar06}.
In the classic reconnection theory by Sweet (1958) and Parker (1957), the reconnection rate $M_{A}$ is given by $R_m^{-1/2}$, which is 
as low as $\sim 10^{-7}$ in the solar corona because the magnetic Reynolds number (Lundquist number) $R_m = LV_A/\eta 
\sim 10^{14}$, where $L (\sim10^{9}$~cm) is the typical size of the flare and $\eta$ ($\sim10^{13}T^{-3/2}\sim 10^4~{\rm cm^2~s^{-1}}$ for $T\sim 10^6$~K) is the magnetic diffusivity for Spitzer resistivity. This value is too small to explain the timescales and the energy release rates of solar flares. 
However, Petschek (1964) predicted that the reconnection rate $M_A$ can be of the order of $0.01$\,-\,$0.1$, 
if the case when slow shocks emanate from a small diffusion region. 
This value is plausible to explain the observed reconnection rate in solar flares.
MHD simulations have shown that localized resistivity can facilitate fast reconnection 
(e.g., Ugai \& Tsuda 1977; Yokoyama \& Shibata 1994),
where fast reconnection is defined as the reconnection whose reconnection rate is close to unity and almost independent of the magnetic Reynolds number, such as Petschek's model. 
Observational support for Petschek's model is given by Tsuneta (1996).

It is believed that localized resistivity may be realized through kinetic effects in the coronal
plasma, such as microscopic instabilities 
and wave-particle interaction. 
Indeed, magnetospheric observation and numerical simulations have shown that 
fast reconnection sets in when the current sheet thickness becomes comparable to the ion inertia length 
or Larmor length \citep{tre01}.
However, such microscopic scales are tiny ($\sim100$~cm) in the solar corona compared with the 
characteristic scale of flares  ($\sim10^9$~cm). How to connect such a huge gap  in the micro 
and the macro scales is one of the fundamental problems in the reconnection physics in solar flares.

To transcend this scale gap, Shibata \& Tanuma (2001) suggested the fractal reconnection model. In this model, plasmoids (regions of magnetically confined plasma) play an important role not only in achieving fast magnetic reconnection but also in linking the small and large scale physics. A current sheet in an antiparallel magnetic 
field is unstable for the tearing instability, and can create plasmoids \citep{FKR, ste83, lou07}. Once they are formed, the 
plasmoids tend to coalesce with each other to form bigger plasmoids \citep{taj87}. When a plasmoid is ejected from the current sheet, 
rapid inflows are locally induced to fill the evacuated region. This leads to the intermittent enhancement of the reconnection rate. In addition, such plasmoids may be created on various scales. 
In this way plasmoids can fill the scale gap between the micro and 
macro scales of reconnection \citep{taj97}.

Many observations report evidence that plasmoids are created in flare current sheets \citep{ohy98,kli00,kar07}. Asai et al. (2004) found several dark downflows (dark voids)  in EUV images that were ejected intermittently during in the decay phase and also in the impulsive and main phases of a flare. Moreover, a nonthermal emission was associated with the motion of each dark void.
Nishizuka et al. (2010) examined multiple plasmoids ejections from a flare using soft X-ray images. They found that the ejected plasmoids were strongly accelerated during multiple peaks in the hard X-ray emission in a flare.
Both studies imply that the reconnection rate was enhanced by plasmoid ejections or downflows.

From the above examples, it can be seen that the observed reconnection rate is important to give a constraint on magnetic reconnection models for solar flares. 
One way to estimate the reconnection rate observationally is to simultaneously detect both reconnection inflow speed and outflow speed.
Because two-dimensional magnetic reconnection theories predict that the reconnection outflow speed $V_{{\rm outflow}} \sim V_{A}$ (e.g. Priest \& Forbes 2000), the reconnection rate $M_{A}$ is of the order of $V_{{\rm inflow}} / V_{\rm outflow}$.

We simultaneously detected both reconnection inflows and outflows for the flare that appeared on 2010 August 18 in EUV wavelengths and use these values to estimate the reconnection rate of the flare.
We also found successive ejections of plasma blobs and their interaction in the current sheet. 
This study indicates a relationship between the energy release process of the flare and the dynamics of the plasma blobs in the current sheet.

\section{Observations of Global Structure of This Flare}
This flare (C4.5 on the {\it GOES} class, see Figure~1(g) and 1(h)) occurred
in the Active Region NOAA 11099 on 2010 August 18, which was
located beyond the northwest limb at the time the flare occurred. 
The EUV images of this flare were
obtained by the Extreme Ultraviolet Imager (EUVI; Howard et al. 2008;
W\"{u}sler et al. 2004) on the {\it Solar Terrestrial Relation
Observatory} spacecraft {\it Ahead} ({\it STEREO-A}; Kaiser et al. 2008)
and the Atmospheric Imaging Assembly (AIA) on the {\it Solar Dynamics
Observatory} (SDO).
In this letter we used 195~{\AA} images of EUVI and six
wavelengths (94~{\AA}, 131~{\AA}, 171~{\AA}, 193~{\AA}, 211~{\AA}, and
335~{\AA}) images of AIA.
The pixel size and the temporal resolution of 195~{\AA} images taken by
EUVI are about 1.6$^{\prime\prime}$ and 150 sec, respectively.  Those of
each wavelength images taken by AIA are about 0.6$^{\prime\prime}$ and
12 sec, respectively.
The 195~{\AA} channel of EUVI contains Fe~{\sc xii} line with the
formation temperature of 1.6~MK and Fe~{\sc xxiv} line with that of
20~MK.  The six narrowband EUV passbands of AIA are sensitive to
different ionization states of iron.  The 171~{\AA} channel contains
Fe~{\sc x} line, formed at 0.6~MK.  The 193~{\AA} channel is similar to
the 195~{\AA} channel of EUVI.  The 211~{\AA} channel contains Fe~{\sc
xiv}, formed at 2~MK.  The 335~{\AA} channel contains Fe~{\sc xvi} line,
formed at 2.5~MK.  The 94~{\AA} channel contains Fe~{\sc x} line and
Fe~{\sc xviii} line, formed at 1~MK and 7~MK, respectively.  In the
131~{\AA} channel Fe~{\sc xxi} line, formed at 11~MK, is dominant in
flaring regions, while it also contains lower temperature lines such as
Fe~{\sc viii} line, formed at 0.4~MK (see O'dwyer et al. 2010 and Reeves
\& Golub 2011).

We investigated the three-dimensional structure of this flare by using
the 195~{\AA} images taken by {\it STEREO-A}/EUVI (which was located
ahead of the earth with the separation angle of 79.877 degrees at
05:00~UT) and 193~{\AA} images by AIA (see Figure~1).  From the EUVI
images, we can obtain the top view of the flare. The typical two ribbon
structure can be seen in Figure~1(e).
From the AIA images, we can obtain the side view of the flare.  We can
discern the vertical structure and the motion because this flare
was located near the west limb (Figure~1(b)).  Before the flare, a
prominence eruption occurred at around 4:40~UT (Figure~1(a) and 1(d)).
Then, intermittent brightening started at 05:07~UT near the position
where the eruption took place.
A sheet structure appeared at 05:10~UT and the bidirectional ejection
began (Figure~1(b) and 1(e)).  Most of the downward
(solar ward) ejections are bright, but we can also see one dark void.
The dark void also appeared below the sheet structure in 193, 94, and
131~{\AA} images.  In this letter we analyze the dynamics of the
bright ejections in EUV, and we call the bright downward plasma motions
in EUV as downflows.
An upward ejection of a bright plasma blob ($\sim$5~arcsec in size) 
was also observed.  
At 05:15~UT, the sheet structure could not be
discerned in all wavelengths images.  After that, post flare loops became prominent 
below the position where the sheet structure was located
(Figure~1(c) and 1(f)).

\section{Simultaneous Observations of Inflow and Outflow and Current
Sheet Dynamics}
Figure~2 shows the closeup images of the reconnection site obtained by
AIA.  The flare loops can be recognized in 94 and 131~{\AA} images, 
indicating  they may be hot ($>$7~MK).  The sheet structure appeared above the hot
loops.  From the sheet structure, the diffusive plasma ejection and a plasma blob
ejections are observed.  We found the flows toward the sheet structure
(inflows) and bidirectional flows from the sheet structure
(outflows).  The inflow speed was measured using 193~{\AA} images
by tracing the motion of threads toward the sheet structure, and the outflow speed using 193~{\AA} by 
tracing the bidirectional ejection along the sheet structure. 

We measure the velocities of these flows by using slits shown in 
the distance-time diagrams perpendicular and parallel to the sheet structure, respectively (Figure~3(a) and (b)).
The horizontal and the vertical axes of Figure~3(c) and 3(d)
are time and the distance along the slits, respectively. The slit NS (N:
North, S: South) and slit EW (E: East, W: West) are for the measurement
of inflow and outflow velocities, respectively.  The solid lines track
the apparent motion of the plasma.  The numbered slopes in
Figure 3(c) and 3(d) show the velocities of the moving structures.  Note
that the bright band in Figure~3(c) ($\sim$40~arcsec from the point S)
corresponds to a part of the sheet structure since the slit NS crosses
the sheet structure.  The dotted line in Figure~3(d) indicates the
height of the edge of the sheet structure at the time when the sheet
structure appeared.

We identified several threads that move toward the sheet structure in the EUV images. To derive the inflow speed of the threads, we used the distance-time diagram shown in Figure 3(c). The movements of the threads are recognized as bright lines in the diagram (marked with black solid lines). These movements started almost simultaneously ($\sim$05:10 UT) with the appearance of the sheet structure. The apparent speed $V_{\rm{pattern}}$ changed from 90~km~s$^{-1}$ to 12~km~s$^{-1}$ during the 5-minute period we could observe the threads.
This indicates that the
inflow velocity decreased several minutes 
after the sheet structure appeared.
If we consider this motion as inflow, this asymptotic value of
velocity ($\sim$12~km~s$^{-1}$) is within the range of other observed
values ($2.6$\,-\,$38$~km~s$^{-1}$) \citep{nar06}.

The speed of the bidirectional flow (outflow) was derived from
Figure~3(d).  The velocities of upward ejections from the sheet
structure were obtained by tracing the plasma ejections and the bright
blob in the 193~{\AA} images (see Figure~2 and
Figure~3(d)). 
Their velocities are $220$\,-\,$250$~km~s$^{-1}$ and
460~km~s$^{-1}$, respectively.  The downflow speed is $250$\,-\,$280$~km~s$^{-1}$.
These ejections are recognized not only in low temperature wavelength images such as 171~{\AA}
but also in high-temperature wavelength images such as 335, 94, and 131~{\AA}.  This
implies that these ejecta contain plasma with a wide temperature
range.  Those velocities are consistent with other observations
\citep{mc99,inn03,asa04}.  
It should be noted that the downflows slowed
down to approximately $70$\,-\,$80$~km~s$^{-1}$ and spread out after colliding
with the loops below.

We also observed many plasma blobs in the sheet structure.  The length
of the sheet is $\sim$20~arcsec and the typical size of these plasma blobs
is $\sim$3~arcsec.
They formed in the sheet, and then are ejected or collide
with each other. Figure~4 shows the time sequence images of 193~{\AA}.  
At first, there was one plasma blob with the size of 2~arcsec in
the sheet structure (Figure~4(a)). Twelve seconds later, its size
became larger ($\sim$3 arcsec) and two plasma blobs (their sizes are
$\sim$2~arcsec) appeared on both sides of the first 
plasma blob (Figure~4(b)). The newly formed plasma blobs increased in
size ($\sim$3~arcsec) in Figure 4(c). After that, they seem to collide
and possibly merge with each other (Figure~4(d)) and are ejected downward at a
speed of $\sim$280~km~s$^{-1}$ (Figure~4(f) and 4(g)).  
They were strongly decelerated when they collided with the loops below, almost stopping. 
The other downward ejections were typically decelerated 
to $70$\,-\,$80$~km~s$^{-1}$ after colliding with the loops below.  
The deceleration can be recognized as the bend of the solid lines 
(which trace the motions of the bright structures) 
in the distance-time diagram shown in Figure~3(d).

\section{Discussion}
We have examined in detail the morphology and dynamics of the the magnetic 
reconnection region in the limb flare on 2010 August 18.  
The overall characteristics, namely large scale eruption, inflow and outflow 
and hot loops below, are consistent with the classical CSHKP model. 
We could examine the fine scale dynamics of the reconnection region 
owing to the high spatial and temporal resolutions of AIA. 
Figure~5 shows the schematic picture of this flaring region.
We observed simultaneous
reconnection inflows and outflows and measured their velocities.  
A lot of plasma blobs appeared in the sheet and
collided with each other or were ejected from it.  
We consider that the sheet structure was the reconnecting current sheet and 
that the plasma blobs were the magnetic islands or plasmoids
created by the tearing instability.
Figure~3(d) is similar to the distance-time diagram obtained from the simulation result of Shen et al. (2011)
(see Figure~6 in their paper) where the plasmoids are ejected bidirectionally, which supports our assertions.
Some plasmoids were made up of the multi-thermal plasma 
because they are recognized in the six wavelength images.
These plasmoids could contain hot ($>$7~MK) plasma as they were visible in the AIA channels sensitive to hot plasma (94 and 131~{\AA}), but these channels are also sensitive to cooler plasma \citep{fos11}, which makes diagnosis of the temperature difficult.
There is a possibility that they were 
heated by the coalescence of plasmoids \citep{kli00}.

We derive the reconnection rate from the observed values by using the relation $M_{A} \sim V_{{\rm inflow}} / V_{\rm outflow}$.
The upward outflow velocity was found to change from $\sim$460~km~s$^{-1}$ to $\sim$220~km~s$^{-1}$, assuming that the plasmoid velocity ejected upward ($\sim$460~km~s$^{-1}$) is almost the same as the outflow velocity.
It should be noted that we did not find another upward ejection after the second ejection shown in Figure~3.
Therefore, we use 460~km~s$^{-1}$ as the outflow velocity in the period of time between 5:10:30 UT and 5:11:00 UT and 220~km~s$^{-1}$ after 5:11:00 UT (see Figure~3).
It should be noted that the apparent
motion in the EUV images is due to the inflow but may not be the true
plasma motion \citep{che04}.  
Spectroscopic observation of
reconnection region is necessary to quantitatively address the flow
velocity \citep{lin05, har06, wan07}. 
We assume that inflow speed was
close to the apparent motion speed.  Because the length of 
the visible edge of the sheet did not change drastically,
we neglect any effects for this motion on the inflow speed \citep{yok01}.
If projection effects are neglected, the
inflow speed can be estimated as $V_{{\rm inflow}} \sim V_{{\rm pattern}}$.
The pattern velocity changed from 90~km~s$^{-1}$ to 12~km~s$^{-1}$, and
therefore, the reconnection rate $M_{\rm A}$ changed from 0.20 to 0.055
within the 5 minute period starting from when the current sheet became bright in EUV.  This
value is comparable with the predicted value from Petschek's (1964) reconnection model  
and other observed values \citep{nar06}.

We found that the disappearance of the plasmoids in the later phase coincides with the reduction of the reconnection rate from 0.20 to 0.055 (see Figure~3 and 4). Theoretically, plasmoid ejections induce strong inflow, which can promote the thinning of the current sheet and enhance the reconnection rate \citep{shi01}.
Therefore the reduction of the reconnection rate may be due to the suppression of the plasmoid activities.
Although we lack detailed information of the time and space variations of the inflow velocity
near the current sheet, which is necessary to confirm the above assertion,
this result probably suggests that the reconnection rate of this flare was enhanced by the plasmoid dynamics.

\acknowledgments
We would like thank Dr. A. Hillier for fruitful comments on the manuscript.  We are
grateful to SDO/AIA \& STEREO/EUVI teams for providing the data used in
this study.  This work was supported by the Grand-in-Aid for the Global
COE Program ``The Next Generation of Physics, Spun from Universality \&
Emergence'' from the Ministry of Education, Culture, Sports, Science and
Technology (MEXT) of Japan.
AA is supported by KAKENHI (23340045).
HI is supported by the Grant-in-Aid for Young Scientists (B, 22740121).



{\it Facilities:} \facility{SDO}, \facility{STEREO}


\clearpage

\begin{figure}
\epsscale{.60}
\plotone{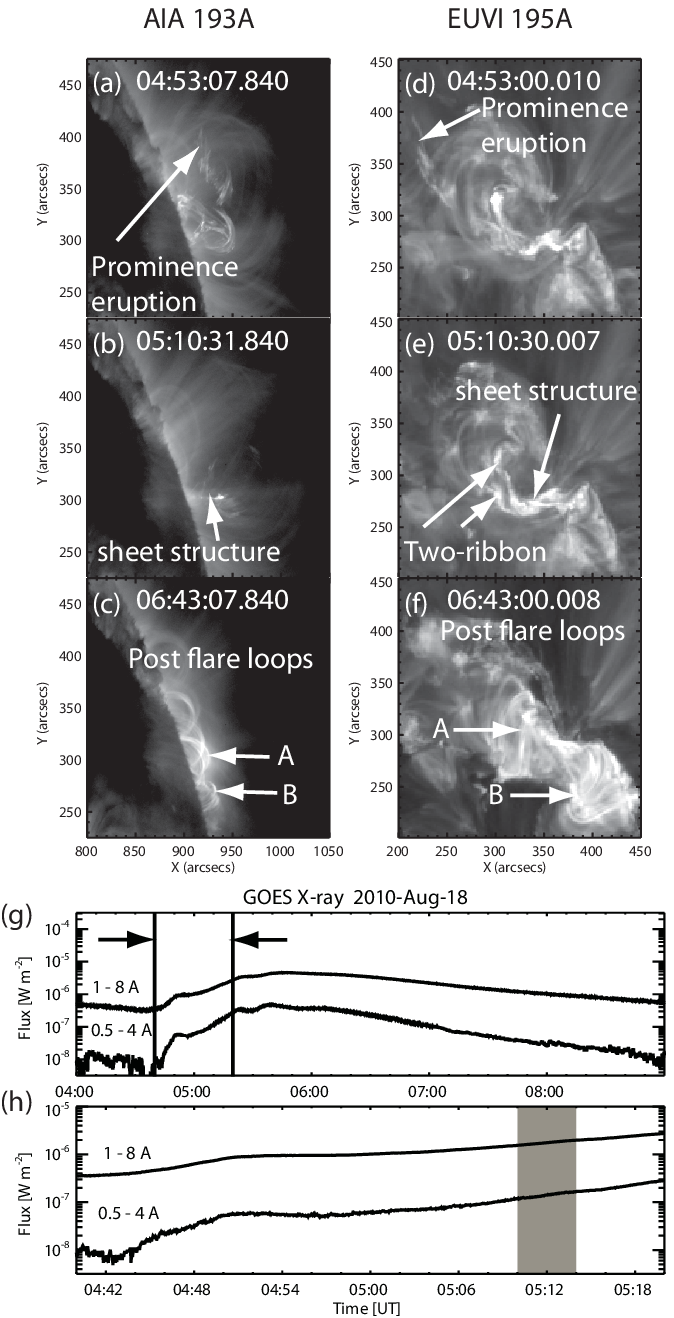}
\caption{Evolution of the 2010 August 18 flare.
{\it Top}: Comparison between images of SDO/AIA 193 {\AA}({\it
Left}: a -- c) and STEREO/EUVI 195~{\AA}({\it Right}: d -- f).
{\it Bottom}: {\it GOES} soft X-ray flux in 0.5 -- 4.0~{\AA} and 1.0 -- 8.0~{\AA}
channel.  The time range of the bottom panel is indicated by arrows in
the top panel.  The gray region in the bottom panel corresponds to the
time span studied in Figure~3.\label{fig1}}
\end{figure}

\begin{figure}
\epsscale{.80}
\plotone{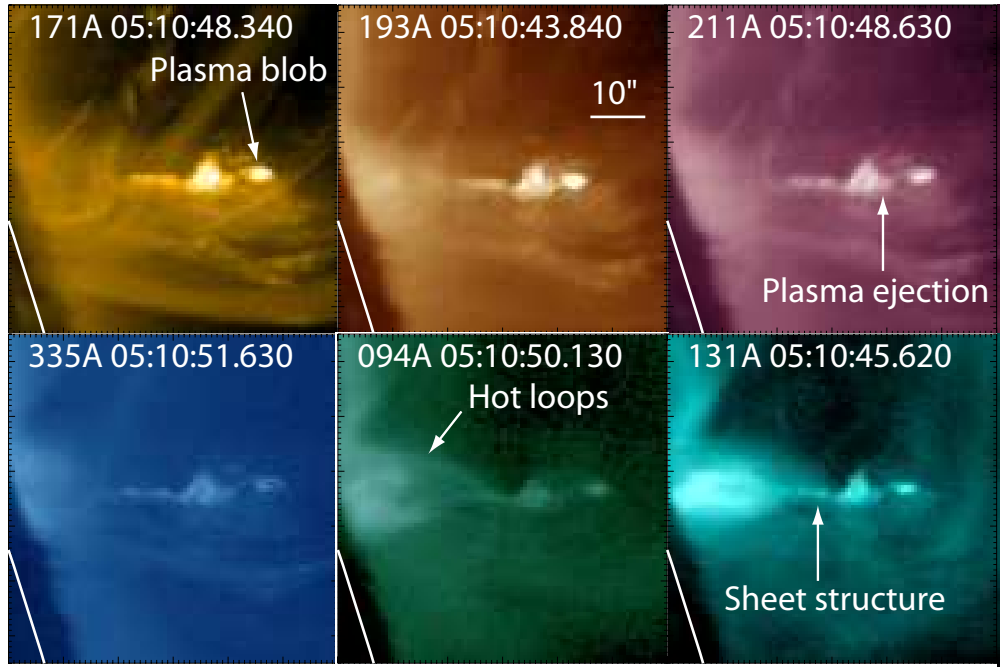}
\caption{Closeup images of the reconnection site in six
different wavelengths (171, 193, 211, 335, 94 and 131 {\AA}) of AIA at
the time when the current sheet, the plasma blob and the hot post flare
loops are observed. White solid lines indicate the solar
limb. \label{fig2}}
\end{figure}

\begin{figure}
\epsscale{.80}
\plotone{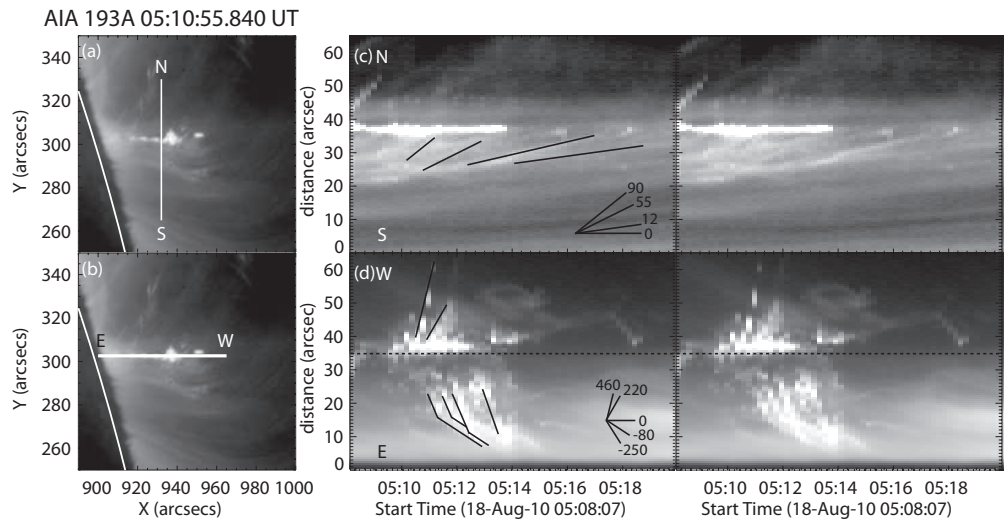}
\caption{{\it Left}: Images of the sheet structure with slits (AIA 193 {\AA}). {\it Right}: Time-sequenced images obtained along the slit NS (N: North, S: South, Fig.3(a)) and the slit EW (E: East, W: West Fig.3(b)). The numbers of each line represent the velocity corresponding to the slope (km $\rm s^{-1}$). The dotted line in the time-sequenced image of slit EW indicates the hight of a visible edge of the sheet structure at the time when the current sheet appeared. \label{fig5}}
\end{figure}

\begin{figure}
\epsscale{.40}
\plotone{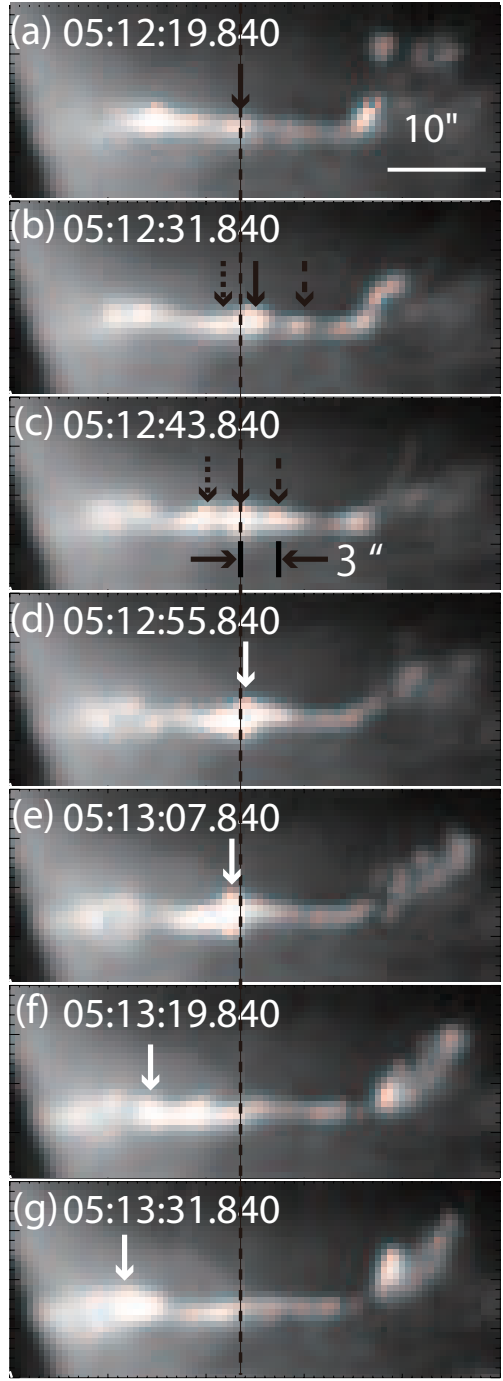}
\caption{Time-sequenced images of the plasma blobs in the current sheet (AIA 193 {\AA}). The arrows (dotted-arrow, solid-arrow and broken-arrow) indicate the positions of the plasma blobs. 
The plasma blobs are marked with their own individual arrow, which follow their evolution.
The white arrows indicate the position of the coalesced plasma blobs. The dashed line indicates the position of the plasma blob pointed by the solid-arrow at 05:12:19.840 UT.}
\end{figure}

\begin{figure}
\epsscale{.60}
\plotone{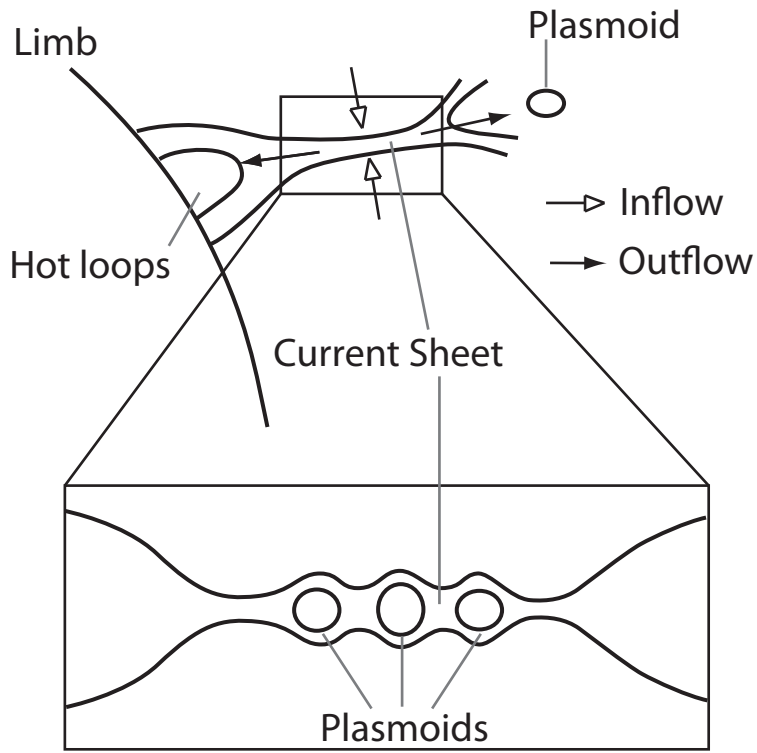}
\caption{A schematic diagram of the flaring region. Black solid lines indicate the magnetic field. {\it Top}: The global configuration of the magnetic field. {\it Bottom}: a closeup image of the current sheet region.}
\end{figure}

\clearpage


\begin{thebibliography}{}
\bibitem[Asai et al.(2004)]{asa04} Asai, A., Yokoyama, T., Shimojo, M., \& Shibata, K. 2004, \apjl,
    605, L77

\bibitem[Carmichael(1964)]{car64} Carmichael, H.\ 1964, NASA 
Special Publication, 50, 451 

\bibitem[Chen et al.(2004)]{che04} Chen, P.~F., Shiabta, K., Brooks, D.~H., \& Isobe, H. 2004, \apjl, 602, L61

\bibitem[Foster \& Testa(2011)]{fos11} Foster, A.~R., \& Testa, P.\ 2011, \apjl, 740, L52 

\bibitem[Furth et al.(1963)]{FKR} Furth, H.~P., Killeen, J., \& Rosenbluth, M.~N.\ 1963, Physics of Fluids, 6, 459

\bibitem[Hara et al.(2006)]{har06} Hara, H., Nishino, Y., Ichimoto, K., \& Delaboudini\'{e}re, J. -P. 2006, \apj, 648, 712

\bibitem[Hirayama(1974)]{hir74} Hirayama, T.\ 1974, \solphys, 34, 323

\bibitem[Howard et al.(2008)]{how08} Howard, R.~A., Moses, J.~D., Vourlidas, A., et al.\ 2008, \ssr, 136, 67

\bibitem[Innes et al.(2003)]{inn03} Innes, D.~E., McKenzie, D.~E., \& Wang, T.\ 2003, \solphys, 217, 247

\bibitem[Isobe et al.(2002)]{iso02} Isobe, H., Yokoyama, T., Shimojo, M., Morimoto, T., Kozu, H., Eto, S., Narukage, N., \& Shibata, K. 2002, \apj, 566, 528
	 
\bibitem[Isobe et al.(2005)]{iso05} Isobe, H., Takasaki, H., \& Shibata, K. 2005, \apj, 632, 1184

\bibitem[Kaiser et al.(2008)]{kai08} Kaiser, M.~L., Kucera, T.~A., Davila, J.~M., et al.\ 2008, \ssr, 136, 5

\bibitem[Karlick{\'y} \& B{\'a}rta(2007)]{kar07} Karlick{\'y}, M., \& B{\'a}rta, M. 2007, \aap, 464, 735

\bibitem[Kopp \& Pneuman(1976)]{kp76} Kopp, R.~A., \& Pneuman, G.~W.\ 1976, \solphys, 50, 85 

\bibitem[Kliem et al.(2000)]{kli00} Kliem, B., Karlick{\'y}, M., \& Benz, A. O. 2000, \aap, 360, 715

\bibitem[Li \& Zhang(2009)]{li09} Li, L., \& Zhang, J.\ 2009, \apj, 703, 877

\bibitem[Lin et al.(2005)]{lin05} Lin, J., Ko, Y. -K., Sui, L., Raymond, J. C., Stenborg, G. A., Jiang, Y., Zhao, S., \& Mancuso, S.
	2005, \apj, 622, 1251

\bibitem[Liu et al.(2010)]{liu10} Liu, R., Lee, J., Wang, T., Stenborg, G., Liu, C., \& Wang, H. 2010, \apjl, 723, L28

\bibitem[Loureiro et al.(2007)]{lou07}Loureiro, N.~F., Schekochihin, A.~A., \& Cowley, S.~C.\ 2007, Physics of Plasmas, 14, 100703 

\bibitem[Masuda et al.(1994)]{mas94} Masuda, S., Kosugi, T., Hara, H., Tsuneta, S., \& Ogawara, Y. 1994,  \nat, 371, 495

\bibitem[McKenzie \& Hudson(1999)]{mc99} McKenzie, D. E., Hudson, H. S. 1999, \apjl, 519, L93

\bibitem[Narukage \& Shibata(2006)]{nar06} Narukage, N., Shibata, K. 2006, \apj, 637, 1122

\bibitem[Nishizuka et al.(2010)]{nis10} Nishizuka, N., Takasaki, H., Asai, A., \& Shibata, K. 2010, \apj, 711, 1062

\bibitem[O'dwyer et al.(2010)]{odw10} O'dwyer, B., Del Zanna, G., Mason, H. E., Weber, M. A.,
	\& Tripathi, D. 2010, \aap, 521, A21

\bibitem[Ohyama \& Shibata(1998)]{ohy98} Ohyama, M. \& Shibata, K. 1998, \apj, 499, 934

\bibitem[Parker(1957)]{par57} Parker, E. N. 1957, \jgr, 62, 509

\bibitem[Petschek(1964)]{pet64} Petschek, H.~E.\ 1964, NASA 
Special Publication, 50, 425

\bibitem[Priest 
\& Forbes(2000)]{pri00} Priest, E., \& Forbes, T.\ 2000, Magnetic reconnection : MHD theory and applications / Eric Priest, Terry Forbes.~ New York : Cambridge University Press, 2000.

\bibitem[Reeves \& Golub(2011)]{ree11} Reeves, K. K., \& Golub, L. 2011, \apjl, 727, L52

\bibitem[Savage 
\& McKenzie(2011)]{sav11} Savage, S.~L., \& McKenzie, D.~E.\ 2011, \apj, 730, 98

\bibitem[Shen et al.(2011)]{she11} Shen, C., Lin, J., \& Murphy, N.~A.\ 2011, \apj, 737, 14

\bibitem[Shibata et al.(1995)]{shi95} Shibata, K., Masuda, S., Shimojo, M., Hara, H., Yokoyama, T., 
Tsuneta, S., Kosugi, T., \& Ogawa, Y. 1995, \apjl, 451, L83

\bibitem[Shibata(1999)]{shi99} Shibata, K. 1999, \apss, 264, 129

\bibitem[Shibata 
\& Tanuma(2001)]{shi01} Shibata, K., \& Tanuma, S.\ 2001, Earth, Planets, and Space, 53, 473

\bibitem[Steinolfson \& Van Hoven(1983)]{ste83} Steinolfson, R. S. \& van Hoven, G. 1983, Physics of Fluids, 26, 117

\bibitem[Sturrock(1966)]{stu66} Sturrock, P. A. 1966, Nature, 211, 69

\bibitem[Sui \& Holman(2003)]{sui03} Sui, L., \& Holman, G.~D.\ 2003, \apjl, 596, L251

\bibitem[Sweet(1958)]{swe58} Sweet, P. A. 1958,  in {\it Electromagnetic Phenomena in Cosmical
Physics}, edited by B. Lehnert (Cambridge University
Press, New York, 1958), pp. 123

\bibitem[Tajima et al.(1987)]{taj87} Tajima, T., Sakai, J., Nakajima, H., Kosugi, T., Brunel, F., \& Kundu, M. R. 1987, \apj, 321,1031

\bibitem[Tajima 
\& Shibata(1997)]{taj97} Tajima, T., \& Shibata, K.\ 1997, Plasma astrophysics (Reading, MA: AddisonWesley)

\bibitem[Treumann(2001)]{tre01} Treumann, R.~A.\ 2001, Earth, Planets, and Space, 53, 453 

\bibitem[Tsuneta et al.(1992)]{tsu92} Tsuneta, S., Hara, H., Shimizu, T., Acton, L. W.,
	 Strong, K. T., Hudson, H. S., \& Ogawara, Y. 1992, \pasj, 44, L63

\bibitem[Tsuneta(1996)]{tsu96} Tsuneta, S.\ 1996, \apj, 464, 1055

\bibitem[Tsuneta(1997)]{tsu97} Tsuneta, S.\ 1997, \apj, 483, 507
	 
\bibitem[Wang et al.(2007)]{wan07} Wang, T., Sui, L., \& Qiu, J. 2007, \apjl, 661, L207

\bibitem[Wuelser et al.(2004)]{wul04} Wuelser, J.-P., Lemen, J.~R., Tarbell, T.~D., et al.\ 2004, \procspie, 5171, 111

\bibitem[Yamada et al.(2010)]{yam10} Yamada, M., Kulsrud, R., 
\& Ji, H.\ 2010, Reviews of Modern Physics, 82, 603 

\bibitem[Yokoyama et al.(2001)]{yok01} Yokoyama,T., Akita, K., Morimoto,T., Inoue, K., 
	\& Newmark, J. 2001, \apjl, 546, L69

\end{thebibliography}
\end{document}